\documentclass[journal]{IEEEtran}

\usepackage{amsmath}
\usepackage{amsfonts}
\usepackage{amssymb}
\usepackage[ruled,vlined]{algorithm2e}

\DeclareMathOperator{\sinc}{sinc}

\usepackage{blindtext, graphicx}

\usepackage[font=scriptsize,caption=false,labelsep=space]{subfig}
\usepackage{graphicx,float,wrapfig}
\usepackage{mathtools}
\usepackage{nicefrac}
\usepackage{bm}
\usepackage{txfonts}
\usepackage{multirow}

\usepackage{verbatim}

\usepackage{url}

\newcommand{\subparagraph}{}

\usepackage{cite}

\usepackage{balance}

\usepackage{color}

\usepackage{upquote}
\usepackage[none]{hyphenat}
\begin{document}

\title{Intelligent Time-Varying Metasurface Transceiver for Index Modulation in 6G Wireless Networks}

\author{John~A.~Hodge,~\IEEEmembership{Student~Member,~IEEE,}
        Kumar~Vijay~Mishra,~\IEEEmembership{Senior~Member,~IEEE,}
        and~Amir~I.~Zaghloul,~\IEEEmembership{Life~Fellow,~IEEE}
\thanks{J. A. H. and A. I. Z. are with Bradley Department of Electrical and Computer Engineering, Virginia Tech, Blacksburg, VA 24061 USA. Email: \{jah70, amirz\}@vt.edu.}
\thanks{K. V. M. and A. I. Z. are with United States CCDC Army Research Laboratory, Adelphi, MD 20783 USA. E-mail: kumarvijay-mishra@uiowa.edu, amirz@vt.edu.}
\thanks{J. A. H. acknowledges support from Northrop Grumman Mission Systems (NGMS), Baltimore, MD, for his thesis research. K. V. M. acknowledges support from the National Academies of Sciences, Engineering, and Medicine via Army Research Laboratory Harry Diamond Distinguished Postdoctoral Fellowship.}
}

\markboth{IEEE ANTENNAS AND WIRELESS PROPAGATION LETTERS, 2020}%
{Shell \MakeLowercase{\textit{et al.}}: Bare Demo of IEEEtran.cls for IEEE Journals}

\maketitle

\begin{abstract}
Index modulation (IM) is one of the candidate technologies for the upcoming sixth generation (6G) wireless communications networks. In this paper, we propose a space-time-modulated reconfigurable intelligent metasurface (RI-MTS) that is configured to implement various frequency-domain IM techniques in a multiple-input multiple-output (MIMO) array configuration. Unlike prior works which mostly analyze signal theory of general RI-MTS IM, we present novel electromagnetics-compliant designs of specific IMs such as sub-carrier index modulation (SIM) and MIMO orthogonal frequency-domain modulation IM (MIMO-OFDM-IM). Our full-wave electromagnetic simulations and analytical computations establish the programmable ability of these transceivers to vary the reflection phase and generate frequency harmonics for IM. Our experiments for bit error rate show that RI-MTS-based SIM and MIMO-OFDM-IM are lower than conventional MIMO-OFDM.
\end{abstract}

\IEEEpeerreviewmaketitle

\section{Introduction}
\IEEEPARstart{T}{ime}-modulated antenna arrays, whose radiated power pattern is steered by varying the width of the periodic pulses applied to each element, are long known to have applications in side-lobe reduction \cite{kummer1963ultra,yang2002sideband}, harmonic beamforming \cite{poli2011harmonic}, and directional modulation in phased arrays \cite{daly2009directional}. Such arrays based on metasurfaces (MTSs) have drawn significant interest in the engineering community \cite{hadad2015space} because of their ability to control and manipulate electromagnetic (EM) waves in a sub-wavelength thickness through modified boundary conditions \cite{chen2016review,glybovski2016metasurfaces}. The MTS, viewed as a two-dimensional (2-D) equivalent of metamaterials, is a synthetic electromagnetic surface composed of sub-wavelength patches, or meta-atoms, printed on one or more dielectric substrate layers \cite{nguyen2019retrieval}. Through careful engineering of each meta-atom, MTSs can transform an incident EM wave into an arbitrarily tailored transmitted or reflected wavefront \cite{hodge2019rf,hodge2019joint,hodge2019multi,hodge2020reflective}.

Recent developments in spatio-temporally (ST) modulated MTSs have unlocked a new class of nonlinear and nonreciprocal behaviors, including direct modulation of carrier waves \cite{hadad2015space}, programmable frequency conversion \cite{salary2018electrically,zhang2018space}, controllable frequency harmonic generation \cite{shaltout2019spatiotemporal}, and cloaking \cite{caloz2019spacetime, shaltout2019spatiotemporal}. These properties are very attractive for designing future low-cost and light-weight wireless communications systems where control of beam-pattern is key to enable reliable and efficient information delivery through massive multiple-input multiple-output (MIMO) antenna arrays \cite{mishra2019reconfigurable}. Notably, reconfigurable intelligent MTSs (RI-MTSs) are capable of applying dynamic transformations of EM waves and have been recently proposed as sensors in fifth/sixth-generation (5G/6G) smart radio environments \cite{di2019smart}. An RI-MTS employs an array of individually-controllable meta-atoms to scatter incident signals to maximize metrics such as receiver signal-to-noise ratio (SNR) \cite{mishra2019reconfigurable}. While several theoretical studies analyze signal processing for 5G/6G RI-MTSs \cite{basar2019wireless,di2019smart,hodge2020coded} and large intelligent surfaces (LISs) \cite{basar2019large}, their specific EM analyses remain unexamined. 

Contrary to these works, we focus on EM analysis and implementation of RI-MTSs for wireless communications. In particular, we consider transceiver for index modulation (IM) that is identified as one of the preferred 5G/6G technologies \cite{mao2018novel} largely because of better energy and spectral efficiencies than conventional modulations \cite{cheng2018index}. The information in IM is encoded through permutations of indices of spatial, frequency, or temporal media. Common IM techniques \cite{ishikawa201850,hodge2020media} include spatial modulation \cite{basar2016index} and subcarrier IM (SIM) \cite{abu2009subcarrier}.

Our prior work \cite{hodge2019reconfigurable} introduced the concept of RIS-based spatial modulation. Motivated by recent research in ST-modulated MTSs, we hereby propose and demonstrate RIS-based designs for a variety of IM techniques such as frequency shift keying (FSK) \cite{bouida2015reconfigurable}, and orthogonal frequency-division multiplexing (OFDM) with IM (OFDM-IM) \cite{basar2013orthogonal,mao2018novel}, and MIMO-OFDM-IM \cite{basar2016multiple}. We implement these frequency-domain IM techniques using the concepts of ST-modulated metamaterials and reflect-array antennas. Our full-wave EM simulations for meta-atom design validate the scattering radiation pattern of our finite RI-MTS array. Finally, we validate the RIS performance using wireless communications model and establish that, despite occupying less spectrum, our proposed designs result in bit error rates (BERs) that are lower than traditional OFDM.

\section{System Model}
\label{sec:im}
Consider a MIMO-OFDM wireless system with $N_t$ transmit and $N_r$ receive antennas. In OFDM, a frequency-selective fading channel is handled by dividing the spectrum into multiple flat-fading subchannels of equal bandwidth \cite{liu2002space}. Unlike standard frequency domain modulation where the carriers are non-overlapping and separated by additional guard bands, the gap between OFDM subcarriers is equal to the inverse of the symbol duration. The resulting overlap of subcarriers, with the peak of one coinciding with the nulls of the other, increases the spectral efficiency \cite{wu1995orthogonal}. Each one of the $N_{s}$ symbols independently modulates one of the equi-bandwidth OFDM subcarriers that are transmitted simultaneously. The sum of the modulated signals is the complex baseband OFDM signal\par\noindent\small
\begin{align}
    x(t) = \frac{1}{\sqrt{N_{s}}} \sum_{a=0}^{N_{s}-1}X_{a}e^{ \mathrm{j} 2\pi at/T_{s}}, 0 \leq t \leq T_{s},
\end{align}\normalsize
where $X_{a}$ are data symbols and $T_{s}$ is symbol duration. The length of the entire message bit sequence is $N_L=MN_s$, where each message sequence vector is a $M$-bit codeword. The transmit signal at the RF stage is ${\bf x}\in \mathbb{C}^{N_{t}\times 1}$.

For line-of-site (LoS) communications, the multipath fading channel follows a Rician distribution \cite{proakis2008digital}. Here, the $N_{r} \times N_{t}$ complex channel impulse response $\mathbf{H}$ is modeled as the sum of the fixed LOS component and a random multipath non-LoS (nLoS) channel component as \cite{paulraj2003introduction}\par\noindent\small
\begin{equation}
    \mathbf{H} = \sqrt{\frac{K}{K+1}} \mathbf{H}_{LoS} + \sqrt{\frac{1}{K+1}} \mathbf{H}_{nLoS},
\end{equation}\normalsize
\noindent where $K$ is the Rician $K$-factor of the channel, $\mathbf{H}_{LoS} \in \mathbb{C}^{N_{r} \times N_{t}}$ is the LoS channel component that is unchanged during the channel coherence time, and $\mathbf{H}_{nLoS} \in \mathbb{C}^{N_{r} \times N_{t}}$ is the nLoS fading component representing random multipath fading. The Rician $K$-factor is the ratio between the power in the direct path (LoS) and the power in the other scattered nLoS paths. Assuming a narrowband block-fading channel, the received signal is \par\noindent\small
\begin{align}
    \mathbf{y}=\mathbf{Hx}+\mathbf{n}, \label{eqn:systemModel}
\end{align}\normalsize
where ${\bf y}\in \mathbb{C}^{N_{r}\times 1}$ is the output of $N_r$ receive antennas  and ${\bf n}\in \mathbb{C}^{N_{r}\times 1}$ is the circularly symmetric white Gaussian noise. The BER is computed after decoding the received symbols usually via maximum likelihood (ML) detector \cite{basar2016index}.

Current MIMO-OFDM techniques have limited energy efficiency caused by power consumption that increases linearly with radio-frequency (RF) chains \cite{elbir2020joint}. The SIM-OFDM or OFDM-IM overcome this \cite{abu2009subcarrier}by employing a new dimension of subcarrier index for modulating additional bits in addition to the usual phase and amplitude indices of the signal constellation. In OFDM-IM, $K_{s}$ of $N_{s}$ subcarrier are activated per symbol leaving $p_{1} = N_{s} - K_{s}$ index bits for signaling. More recently, OFDM-IM has been combined with a MIMO configuration to yield MIMO-OFDM-IM which achieves significantly lower BER than the traditional MIMO-OFDM \cite{basar2016multiple}. The number of bits per channel use (bpcu) for MIMO-OFDM-IM is $N_{t}(p_{1}+\log(M)K_{s})$. Note that when $M=1$ in OFDM-IM, the system is equivalent to FSK \cite{ishikawa201850}.
\begin{figure}[t]
  \centering
  \includegraphics[width=0.95\columnwidth]{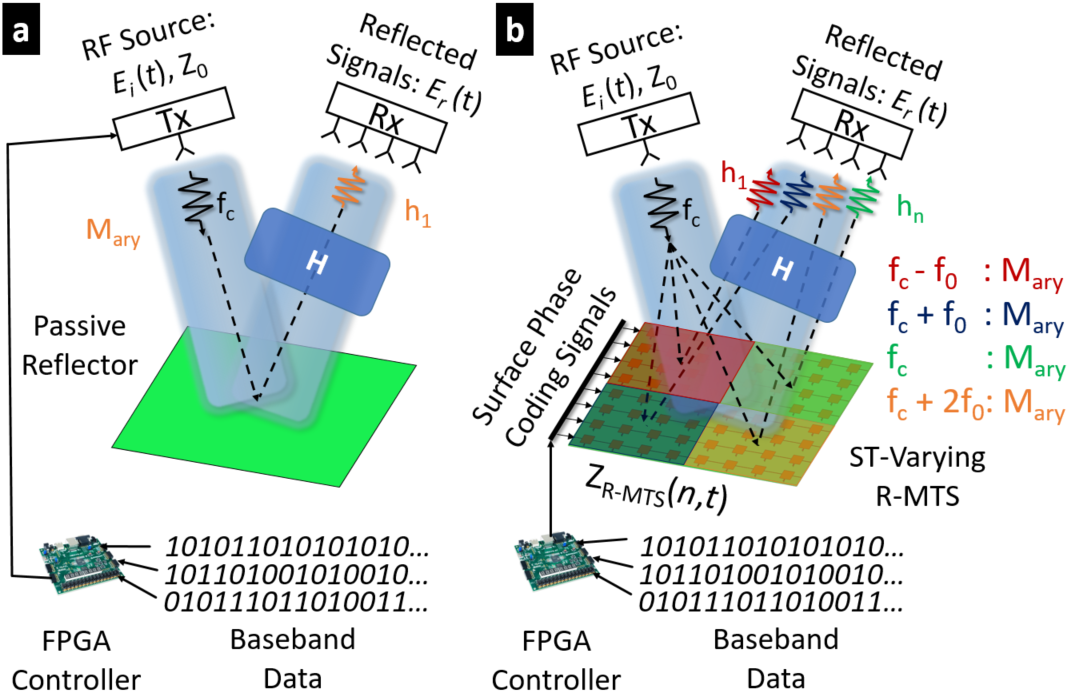}
  \caption{(a) Simplified illustration of a conventional $M$-ary phase shift keying (PSK) system using a passive reflector. (b) A digitally programmable time-varying RI-MTS to implement OFDM-IM. The reflected carrier wave is modulated by varying the complex impedance of the RI-MTS over space and time (ST). By programming the phase coding of the , reconfigurable radiation patterns are tailored to perform frequency translation. The  aperture is comprised of $N_1$ by $N_2$ meta-atoms.}
  \label{fig:SystemGraphic}
\end{figure}

In a conventional wireless communication transceiver with a passive reflector (Fig.~\ref{fig:SystemGraphic}a), modulation is performed using complex RF circuitry in the feed usually only in a single input/output (SI/SO) configuration. For MIMO-based IM, we consider ST-coded RI-MTS (Fig.~\ref{fig:SystemGraphic}b) comprising an array of $N_1\times N_2$ meta-atoms, each of which is embedded with a varactor diode and variable resistor to control the time-varying reflection amplitude and phase of each meta-atom. A field programmable gate array (FPGA) controller and multiplexer translates information bits to a ST coding matrix \cite{hodge2019reconfigurable}.

We model MTS using generalized sheet transition conditions to relate electric field vector $\boldsymbol{\vec{E}} = [E_{x}\ E_{y}\ E_{z}]^T$, where $(\cdot)^T$ denotes the transpose, on both sides of the MTS via surface electric susceptibility tensors $\overline{\overline{\chi }}^{E} =\bigg[\begin{smallmatrix} \chi^{E}_{xx} & \chi^{E}_{xy} & \chi^{E}_{xz} \\ \chi^{E}_{yx} & \chi^{E}_{yy} & \chi^{E}_{yz} \\ \chi^{E}_{zx}  & \chi^{E}_{zy} & \chi^{E}_{zz} \end{smallmatrix}$\bigg] in the Cartesian coordinate system (magnetic field $\boldsymbol{\vec{H}}$ and susceptibility $\overline{\overline{\chi }}^{M}$ are defined similarly): \par\noindent\small
    \begin{align}
    \hat{z}\times \boldsymbol{\vec{H}}\vert _{z=0^{-}}^{0^{+}}&=j\omega \overline{\overline{\chi }}^{E}\cdot \boldsymbol{\vec{E}}_{t,av}\vert _{z=0}-\hat{z}\times\nabla_{t}\left[\chi^{M}_{zz}H_{z,av}\right]_{z=0},\\ \boldsymbol{\vec{E}}\vert _{z=0^{-}}^{0^{+}}\times \hat{z}&=j\omega \overline{\overline{\chi }}^{M}\cdot \boldsymbol{\vec{H}}_{t,av}\vert _{z=0}+\hat{z}\times\nabla_{t}\left[\chi^{E}_{zz}E_{z,av}\right]_{z=0},
    \end{align}\normalsize
where $\hat{z}$ is the unit vector, $\omega$ is the angular frequency, $\nabla$ is the gradient operator, subscript ``$av$'' represents average of fields on either side of MTS, and subscript $t$ refers to tangential field components which are transverse to $z$ \cite{holloway2012overview}. Our reflect-array MTS is fully reflective with no transmit fields. At the surface of the RI-MTS $(z=0)$, the reflected electric field $\boldsymbol{\vec{E}}_{r}$ is \par\noindent\small
\begin{align}
    \boldsymbol{\vec{E}}_{r}(z=0,t) = \Gamma(t) \boldsymbol{\vec{E}}_{i}(z=0,t), \label{eqn:A4}
\end{align}\normalsize
where $\boldsymbol{\vec{E}}_{i}$ is incident field and $\Gamma(t)$ is time-varying complex reflection coefficient. Note that $\Gamma(t) = \frac{ Z_{s}(t) - Z_{0} }{ Z_{s}(t) + Z_{0} }$, where the complex surface impedance $Z_{s}$ characterises the behavior of each meta-atom and $Z_{0} = 377$ $\Omega$ is free-space impedance.

To perform OFDM-IM, Fourier transform of \eqref{eqn:A4} yields the angular frequency response \cite{ramaccia2019phase}\par\noindent\small
\begin{align}
    \boldsymbol{\vec{E}}_{r}(\omega) = \Gamma(\omega)*\boldsymbol{\vec{E}}_{i}(\omega) = \int\Gamma(\omega-\omega')\boldsymbol{\vec{E}}_{i}(\omega')d\omega',  \label{eqn:A5}
\end{align}\normalsize
where $*$ denotes convolution. Given an incident wave of $\boldsymbol{\vec{E}}_{i}(\omega)$, the spectrum of the scattered wave $\boldsymbol{\vec{E}}_{r}(\omega)$ is controlled by varying $\Gamma(\omega)$ of each meta-atom. If $\Gamma(t)$ is a periodic signal, then it is a linear combination of complex exponentials \cite{dai2018independent}, i.e., $\Gamma(t) = \sum_{m=-\infty}^{\infty} a_{m} e^{\mathrm{j}m\omega_{0}t} = \sum_{m=-\infty}^{\infty} a_{m} e^{\mathrm{j}\frac{2 \pi m t}{T_0}}$, where $T_{0}$ is the modulation period, $\omega_{0} = \frac{2 \pi}{T_{0}}$, and $a_{m}$ is the Fourier series (FS) coefficient of the $m$-th harmonic. The corresponding reflected wave in the spectral-domain is\par\noindent\small
\begin{equation}
    \boldsymbol{\vec{E}}_{r}(\omega) = 2 \pi \sum_{m=-\infty}^{\infty} a_{m} \boldsymbol{\vec{E}}_{r}(\omega - m \omega_{0}).
    \label{eqn:ErSpecDomain}
\end{equation}\normalsize
We ST-modulate $\Gamma$ for each meta-atom to radiate the $m$th harmonic frequency or a combination of harmonic frequencies to the desired steer angle. We now propose our meta-atom design and determine its steady-state $\Gamma(\omega)$ at each coding state.

\section{Meta-Atom Design for IM Transceivers}
\label{sec:metaAtomDesign}
Consider a MTS whose $N_1\times N_2$ meta-atoms are indexed by integers $p$ and $q$ along $x$- and $y$-axes, respectively. The inter-element spacing in the $x$ ($y$) dimension is $d_{x}$ ($d_{y}$). Given time-varying complex reflection coefficient $\Gamma_{pq}^{t}$ of the $pq$-th meta-atom, the approximate far-field pattern reflected by a ST-modulated MTS that is illuminated by a plane wave at time $t$ is \cite{zhang2018space,yang2016programmable}\par\noindent\small

\begin{align}
    f(\theta,\phi,t) = \sum_{p=1}^{N_{1}}\sum_{q=1}^{N_{2}}E_{pq}(\theta,\phi) \Gamma_{pq}^{t}
    e^{\mathrm{j} ( k_{c} ((p-1)d_{x}\sin\theta\cos\phi + (q-1) d_{y} \sin\theta\sin\phi) )},  \label{eqn:f101}
\end{align}\normalsize
where $\theta$ and $\phi$ denote angles in spherical coordinates, $E_{pq}(\theta,\phi)$ is the pattern response, and $k_{c} = \frac{2\pi}{\lambda_{c}}$ is the wavenumber. 

We scan the reflected beam by coding the RI-MTS with a progressively varying phase shift. The phase of each meta-atom $\Gamma_{s}(p,q)$ is \cite{huang2007reflectarray}\par\noindent\small
\begin{align}
    \Gamma_{s}(p,q) = e^{-\mathrm{j}k((p-1)d_{x}\sin(\theta_{s})\cos(\phi_{s}) +  (q-1)d_{y}\sin(\theta_{s})\sin(\phi_{s}))},
    \label{eqn:beamScan}
\end{align}\normalsize
where $\{\theta_{s}$, $\phi_{s}\}$ is the desired steering direction of the radiation pattern. The meta-atom response is simulated in HFSS as a unit-cell with periodic boundary conditions. Therefore, we consider local or quasi-periodicity so that the reflection phase is smoothly varying across the MTS \cite{huang2007reflectarray}. From FS, the $m$-th harmonic $a_{pq}^{m}$ of $\Gamma_{pq}(t)$ is \cite{zhang2018space}\par\noindent\small
\begin{equation}
    a_{pq}^{m} = \sum_{n=1}^{L}\frac{\Gamma_{pq}^{n}}{L} \sinc \Big(\frac{\pi m}{L}\Big) e^{-\mathrm{j}\frac{\pi m (2n-1)}{L}}, \label{eqn:f100}
\end{equation}\normalsize
where $L$ is the number of time-steps in the coding sequence per modulation period $T_0$. This provides the amplitude and phase of the $m$th frequency harmonic reflected from the RI-MTS as a function of $\Gamma_{pq}^{n}$. By varying the slope of the phase coding sequence according to $m$, we shift $f_{c}$ to $f_{c} + m f_{0}$ \cite{cumming1957serrodyne, wu2019serrodyne}. To implement OFDM or OFDM-IM, multiple simultaneous sub-carriers are superimposed to produce\par\noindent\small
\begin{equation}
    \Gamma_{pq}^{n} = \sum_{m} X_{m} e^{\mathrm{j}\frac{2\pi m(n-1)}{L} },
    \label{eqn:multHarmonics}
\end{equation}\normalsize
where $X_{m}$ encapsulates the modulated amplitude and phase of each sub-carrier $m$. 

A time shift $t_{q}$ in the periodic time-varying coding sequence is equivalent to spatial phase shift \cite{zhang2018space}, i.e. $\Gamma_{pq}(t - t_q)\mathop{\longleftrightarrow}\limits^{{FS}}a_{pq}^m e^{-\mathrm{j}2\pi mf_0t_q }$. By exploiting this property, we generate the desired IM frequency subcarriers by phase (and/or amplitude) modulation of each meta-atom. As a result, the far-field radiation pattern of the ST RI-MTS at the $m$th harmonic frequency $f_{c} + m f_{0}$ becomes \par\noindent\small
\begin{align}
    f_{m}(\theta,\phi) = \sum_{p=1}^{N_{1}}\sum_{q=1}^{N_{2}}E_{pq}(\theta,\phi) a_{pq}^{m} e^{\mathrm{j} (\frac{2\pi}{\lambda_{c}} ((p-1)d_{x}\sin\theta\cos\phi + (q-1) d_{y} \sin\theta\sin\phi ) )}.  \label{eqn:f102}
\end{align}\normalsize
We use this relation to generate and spatially steer frequency harmonics in a controllable manner for advanced RIS-based modulation, multiplexing, and beamforming.

\begin{figure}[t]
  \centering
  \includegraphics[width=0.90\columnwidth]{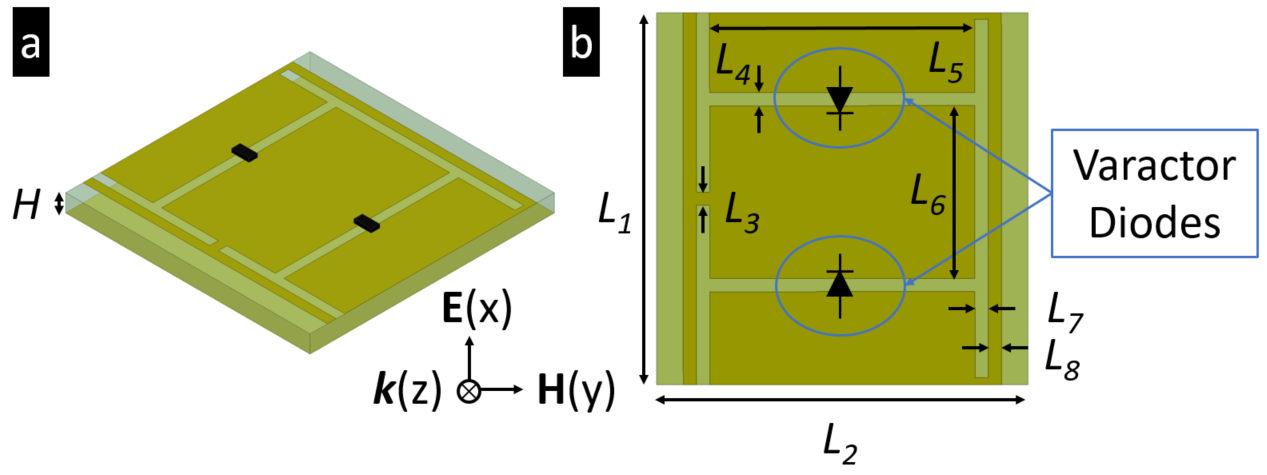}
  \caption{Illustration of proposed time-varying RI-MTS meta-atom. (a) Isometric and (b) Top-down view of the unit cell. Here, $H$ = 0.2 mm, $L_{1}$ = 2.8 mm ($\lambda_{c}/3.83$), $L_{2}$ = 2.8 mm ($\lambda_{0}/3.83$), $L_{3}$ = 0.1 mm, $L_{4}$ = 0.1 mm, $L_{5}$ = 2.0 mm, $L_{6}$ = 1.3 mm, $L_{7}$ = 0.1 mm, and $L_{8}$ = 0.1 mm. The metal traces (top layer) and ground-plane (bottom layer) are copper. The dielectric layer of thickness $H$ is a RT/Duroid{\textregistered} 5880 ($\epsilon_{r} = 2.2$, $\tan \delta = 0.0009 $) substrate. This design has been adapted from \cite{wu2019serrodyne} for $f_{c}=28$ GHz.}
  \label{fig:unitCellDiagramAndDims_v2}
\end{figure}
\begin{figure}[t]
  \centering
  \includegraphics[width=0.90\columnwidth]{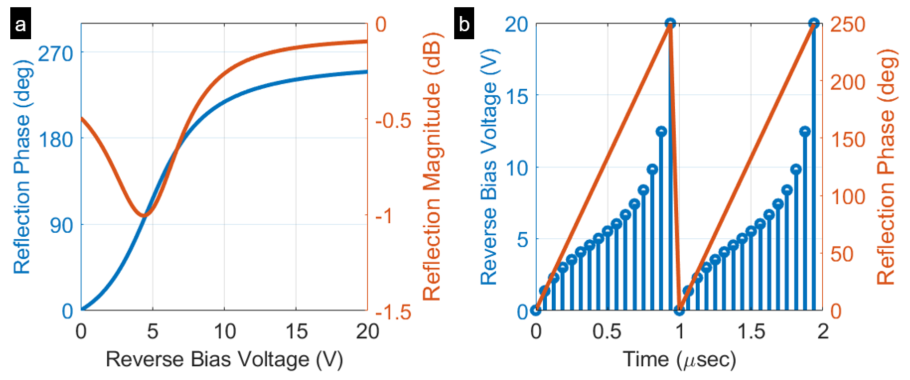}
  \caption{(a) Simulated tunable reflection amplitude and phase response of the meta-atom with embedded varactor diodes at $f_{c} = 28$ GHz (Fig.~\ref{fig:unitCellDiagramAndDims_v2}) using ADS circuit co-simulation with HFSS. (b) Time-harmonic voltage waveform to generate sawtooth linear phase with a period $T_{0} = 1$ $\mu s$.}
  \label{fig:tunable_phase}
\end{figure}
\begin{figure*}[t]
\begin{center}
\includegraphics[width=1.0\textwidth]{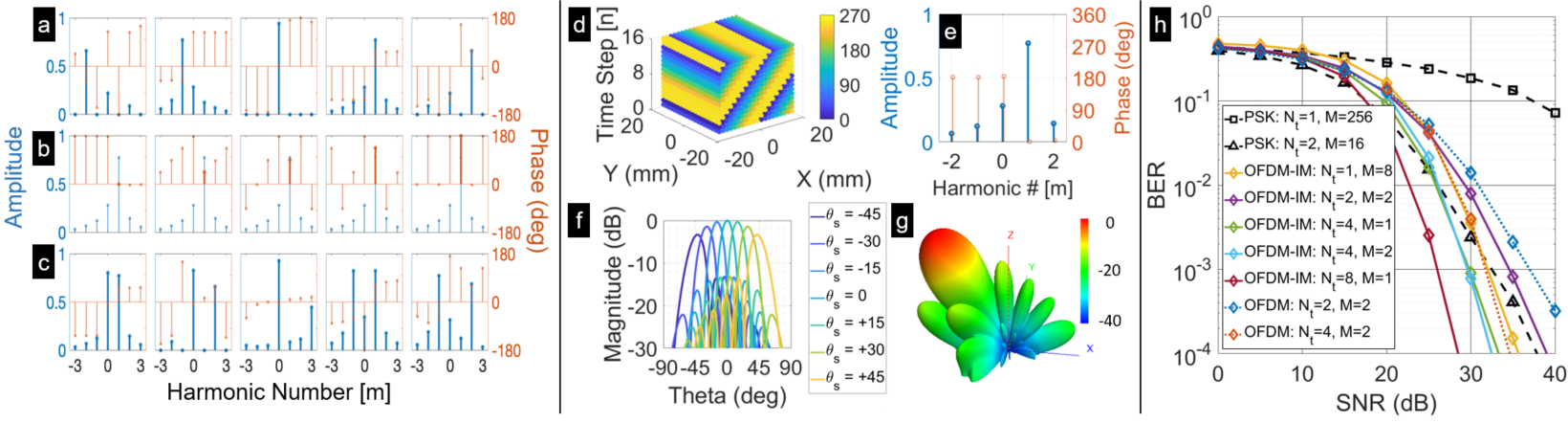}
  \caption{Amplitudes (blue) and phases (red) of $a^{m}_{pq}$ for different time-domain $\Gamma^{n}_{pq}$ coding sequences. (a) Individual frequency harmonics from $m = -2$ to $m = +2$ generated by varying the slope of the meta-atom's time-varying phase as in \eqref{eqn:multHarmonics}. (b) By circularly shifting the time-varying phase coding sequence, $m=+1$ frequency harmonic phase (bold) is phase-modulated to (left to right) $0^{\circ}$, $+45^{\circ}$, $+90^{\circ}$,  $+135^{\circ}$, and $+180^{\circ}$. (c) Examples of multiple simultaneous frequency harmonics generated as per \eqref{eqn:multHarmonics}. (d) The ST coding matrix used to generate $m = +1$ harmonic steered to $\{\theta_{s} = -30^{\circ}, \phi_{s} = 0^{\circ}\}$ at 28 GHz. The color of each dot represents the phase of $\Gamma^{n}_{pq}$ in degrees. (e) Amplitude (blue) and phase (red) of $a^{m}_{pq}$ for $\Gamma^{n}_{11}$ for $n = 1$ to $16$. (f) Normalized radiation pattern ($\phi=0$ cut) scanned from $-45^{\circ}$ to $+45^{\circ}$ in $15^{\circ}$ increments. Only the activated $m = +1$ harmonic is at the peak magnitude. (g) Full-wave finite RI-MTS array scattering simulation of the scanned ($\theta_{s} = -30$ deg) reflected beam. The excitation is a plane wave traveling in the $k=-z$ direction with E-field polarized in the $x$ direction. (h) BER versus SNR performance of various SISO/MIMO-OFDM-IM, a classical MIMO-OFDM (dotted lines), and classical SISO/MIMO-PSK (dashed lines). When $M=1$ (yellow diamond), the system is equivalent to FSK. The OFDM-IM ($N_{t}=8$, $M=1$), OFDM-IM ($N_{t}=4$, $M=2$), and traditional OFDM ($N_{t}=4$, $M=2$) cases are 16 bpcu. All other cases are 8 bpcu.}
  \label{fig:HarmonicGenFig}
\end{center}
\end{figure*}
We designed a reflective time-varying MTS (Fig.~\ref{fig:unitCellDiagramAndDims_v2}) where each meta-atom unit cell is embedded with two varactor diodes. Figure~\ref{fig:unitCellDiagramAndDims_v2}a shows an isometric view of the meta-atom unit cell. To reduce fabrication cost and control complexity, this prototype meta-atom design does not require vias and supports only column-level voltage control. As a result, only 1-D beam scanning is supported in this design. For 2-D beam control, an element-level voltage variation is required to individually address meta-atoms. Among prior works, the closest to our proposed meta-atom is \cite{wu2019serrodyne}. However, our design is modified for reflective  rather than transmit MTS-based phase shifters of \cite{wu2019serrodyne}. We use embedded MACOM MAVR-000120-14110P gallium arsenide (GaAs) flip chip hyperabrupt varactor diodes \cite{macom2020Diode} with tunable capacitance and resistance in the ranges  $0.16$-$1.23$ pF and $0.02$-$0.68$ $\Omega$, respectively. The breakdown voltage for the diode is $20$ V, allowing a tuning range of $0$-$20$ V. The nonlinear Keysight ADS (Advanced Design Systems) diode model that includes losses and capacitance versus reverse bias voltage relationship is available from the manufacturer \cite{macom2020Diode}. Using the S-parameters of the meta-atom simulated in HFSS, we use the ADS diode circuit model to simulate the RF performance of the time-varying meta-atom. We perform ADS harmonic balance simulations to predict $\Gamma(\omega)$ response of MTS modulated with a time-varying voltage signal.

We selected the dimensions of the meta-atom (Fig.~\ref{fig:unitCellDiagramAndDims_v2}b) to provide an operational reflection phase tuning range of $>250^{\circ}$ at $f_{c}$ = 28 GHz, where the meta-atom is relatively low-loss with a reflection amplitude of $<1.0$ dB for all simulated tuning states. Figure~\ref{fig:tunable_phase} shows the reflection amplitude and phase response as a function of diode reverse bias voltage. We synthesized the dimensions of the meta-atom unit cell through parametric tuning and trial-and-error iteration in ANSYS HFSS full-wave EM simulations. The tunable reflection phase range and reflection amplitude loss are primarily functions of diode's capacitance tuning range and its resistance, respectively.
\section{Numerical Experiments}
\label{sec:numexp}

We validated our approach and design through numerical simulations. We nominally considered a $20 \times 20$ meta-atom  with element spacing of $d_{x}=d_{y} = \lambda_{c}/6$ at $f_{c} = 28$ GHz. The meta-atoms are arranged in columns on a rectangular lattice. Control signals independently address the top and/or bottom half of each column to spatially steer the reflected beam to the desired receiver or create multiple sub-apertures for MIMO and spatial modulation techniques. A larger aperture provides more antenna gain for the transmitter, which results in higher SNR at the receiver, or a greater number of sub-apertures for higher communications capacity.

A time-harmonic phase coding sequence with period $T_{0} = 1$ $\mathrm{\mu}$s is used to provide $f_{0} = 1$ MHz harmonic spacing. By specifying the complex weights $X_{m}$ in \eqref{eqn:multHarmonics}, we control the amplitude and phase of each harmonic (Fig.~\ref{fig:HarmonicGenFig}a-c). Figure~\ref{fig:HarmonicGenFig}d-g demonstrate the $m = +1$ harmonic generated and steered to the direction $\{\theta_{s} = -45^{\circ}, \phi_{s} = 0^{\circ}\}$. In particular, Fig.~\ref{fig:HarmonicGenFig}d depicts the corresponding ST coding matrix. The resulting spectrum of $a^{m}_{pq}$ for $\Gamma^{n}_{11}$ for $n = 1$ to $16$ is shown in Fig.~\ref{fig:HarmonicGenFig}e. The scanned radiation patterns of the $m = +1$ harmonic are shown in Fig.~\ref{fig:HarmonicGenFig}f. The full-wave simulation verification of finite RI-MTS array scattered radiation pattern at $\theta_{s} = -30^{\circ}$ is plotted in Fig.~\ref{fig:HarmonicGenFig}g. 

We verified the communications performance of our design through BER performance for RI-MTS-based single-input single-output (SISO) and MIMO-OFDM-IM. In the signal model of \eqref{eqn:systemModel} for OFDM-IM, we set 2 out of $N_{s} = 4$ subcarriers and $p_{1} = 2$ index bits. We benchmark our design against conventional OFDM and PSK systems in terms of their bpcu. Using only $2$ out of $4$ available subcarriers, this OFDM-IM system achieves a $50\%$ reduction in instantaneous bandwidth compared to traditional OFDM. The number of transmit antennas were the same as the number of receive antennas for each case ($N_{t}=N_{r}$). We divided the RI-MTS into 2 and 4 sub-apertures for MIMO operation. We set $K = 10$ dB to be consistent with the experimental characterization at $f_c=28$ GHz \cite{samimi201628} and previous analyses of other LIS-based wireless systems \cite{han2019large}. The BER performance of ML symbol detector (Fig.~\ref{fig:HarmonicGenFig}h) shows that, for same $N_t$, $M$, and number of bits per channel use (bpcu), MIMO-OFDM-IM (purple and green; light blue and red curves) outperforms conventional MIMO-OFDM (dark blue; orange curves, respectively). Note that MIMO-OFDM-IM achieves this BER with less spectrum usage.
\section{Summary}
\label{sec:summ}
We modeled and demonstrated an ST-modulated RI-MTS to perform key IM schemes for 5G/6G wireless networks. Our RIS-based implementations do not require conventional phase shifters, mixers, or a lossy RF manifold network, but need a control network to individually address the antenna elements. Consequently, these RI-MTS architectures hold the promise to achieve direct frequency modulation and beam scanning in significantly less size, weight, and power consumption compared to conventional phased arrays.

\clearpage
\balance
\bibliographystyle{IEEEtran}
\bibliography{main}

\end{document}